\documentclass[aps,prd,12pt,nofootinbib,superscriptaddress]{revtex4-2}
\usepackage{bm}
\usepackage{amsmath}
\usepackage{ulem}
\usepackage{comment}
\usepackage{latexsym}
\usepackage{amsfonts}
\usepackage{graphicx}
\usepackage{mathrsfs}
\usepackage{CJKutf8}
\usepackage{subfigure}
\usepackage{float}
\usepackage{multirow}
\usepackage{orcidlink}
\usepackage{color}
\usepackage{booktabs}
\usepackage{hyperref}
\hypersetup{
	colorlinks=true,
	linkcolor=red,
	citecolor=blue,
}

\begin{document}

\begin{CJK*}{UTF8}{gbsn}

\title{Supernova calibration by gravitational waves}

\author{Xuchen Lu (路旭晨)\,\orcidlink{0000-0002-9093-9059}}
\email{Luxc@hust.edu.cn}
\affiliation{School of Physics, Huazhong University of Science and Technology, Wuhan, Hubei
430074, China}

\author{Yungui Gong (龚云贵)\,\orcidlink{0000-0001-5065-2259}}
\email{Corresponding author. yggong@hust.edu.cn}
\affiliation{School of Physics, Huazhong University of Science and Technology,
Wuhan, Hubei 430074, China}

\begin{abstract}
Hubble tension is one of the most important problems in cosmology.
Although the local measurements on the Hubble constant with Type Ia supernovae (SNe Ia) are independent of cosmological models,
they suffer the problem of zero-point calibration of the luminosity distance.
The observations of gravitational waves (GWs) with space-based GW detectors can measure the luminosity distance of the GW source with high precision.
By assuming that massive binary black hole mergers and SNe Ia occur in the same host galaxy,
we study the possibility of re-calibrating the luminosity distances of SNe Ia by GWs.
Then we use low-redshift re-calibrated SNe Ia to determine the local Hubble constant.
We find that we need at least 7 SNe Ia with their luminosity distances re-calibrated by GWs to reach a 2\% precision of the local Hubble constant.
The value of the local Hubble constant is free from the problems of zero-point calibration and model dependence,
so the result can shed light on the Hubble tension.
\end{abstract}

\maketitle
\section{Introduction}

The value of the Hubble constant is crucial for us to understand the evolution of the Universe
because it characterizes the current expansion rate of the Universe.
Over the years, the measurement precision of the Hubble constant has been drastically improved \cite{Riess:2009pu,Riess:2011yx,Efstathiou:2013via,Riess:2016jrr,Riess:2019cxk,Riess:2020fzl,Riess:2022mme,Freedman:2021ahq,Suyu:2016qxx,Wong:2019kwg,Birrer:2021use,Jimenez:2019onw,DAmico:2019fhj,Ivanov:2019pdj,Colas:2019ret,Birrer:2019otx,Camarena:2019moy,WMAP:2012nax,Planck:2018vyg,Dainotti:2022bzg,Dainotti:2021pqg}.
By recalibrating the extragalactic distance ladder using a sample of Milky Way Cepheids
with the Hubble Space Telescope photometry and Gaia EDR3 parallaxes,
the SH0ES (Supernovae and $H_0$ for the equation of state) team determined the local Hubble constant from Type Ia supernovae (SNe Ia) data as  $H_0=73.15\pm 0.97$ km/s/Mpc \cite{Riess:2022mme}.
Applying the tip of the red giant branch method to SNe Ia data from Carnegie Supernova Project results the Hubble constant,
$H_0=69.8\pm 0.6$ (stat) $\pm 1.6$ (sys)km/s/Mpc \cite{Freedman:2021ahq}.
Combining the strong lensing time delay data and type Ia supernova (SN Ia) luminosity distances,
it was found that $H_0=74.2^{+3.0}_{-2.9}$ km/s/Mpc \cite{Collett:2019hrr}.
However, the measurements of the anisotropies in the cosmic microwave background (CMB) by Planck 2018 based on the $\Lambda$CDM model gave $H_0=67.4\pm 0.5$ km/s/Mpc \cite{Planck:2018vyg}.
These results showed that the values of the Hubble constant determined from different observations are in discrepancy
and suggested that the local measurements and the values inferred from CMB are in significant tension \cite{DiValentino:2021izs}.
As the measurement precision improves, the tension becomes more significant,
we are at a crossroads \cite{Freedman:2017yms}.
As discussed above, the results from the early Universe probe of CMB depend on the $\Lambda$CDM model.
The local measurements from SN Ia standard candles are independent of cosmological models,
but they suffer the zero-point calibration problem due to the uncertainties of the absolute calibration of the peak luminosity for SN Ia and the determination of the absolute distance scale for the luminosity distances.
Furthermore, if we consider the dependence of intrinsic luminosity on color and redshift, the measured value of the Hubble constant changes \cite{Tutusaus:2018ulu,Mortsell:2021tcx}.

The observations of gravitational waves (GWs) can measure the luminosity distance of the GW source with high precision,
providing an independent method of measuring cosmological distances.
In 1986, Schutz proposed to determine the Hubble constant with GWs from binary neutron stars (BNS) \cite{Schutz:1986gp}.
If electromagnetic counterparts of the coalescence of massive binary black hole (MBBH) or BNS can be identified, then the redshift of the GW source is determined
and the luminosity-redshift relation provided by GWs as standard sirens \cite{Holz:2005df} can be used to study the evolution of the Universe \cite{Seto:2001qf,deSouza:2021xtg,Klein:2015hvg,Namikawa:2015prh,Gray:2019ksv,Wolf:2019hun,Farr:2019twy,Keeley:2019hmw,Finke:2021aom}.
In addition to being standard sirens, the propagation of GWs can also probe the evolution of the Universe \cite{Alfaro:2019sbq,Liu:2019dds}. 
Since the first direct observation of GWs by the Laser Interferometer Gravitational-Wave Observatory (LIGO) Scientific Collaboration and the Virgo Collaboration in 2015, there have been reported tens of GW detections
\cite{LIGOScientific:2016aoc,LIGOScientific:2018mvr,LIGOScientific:2020ibl,LIGOScientific:2021usb,LIGOScientific:2021djp}.
The first observed BNS merger GW170817 and its counterpart GRB 170817A gives $H_0=70.0^{+12.0}_{-8.0}$ km/s/Mpc \cite{LIGOScientific:2017adf}.
In the absence of a counterpart one can employ statistical methods, 
by establishing a correlation between GW source and its potential galaxy catalog, to get the redshift of GW source.
Applying this method to 47 GWs from the Third LIGO-Virgo-KAGRA Gravitational-Wave Transient Catalog (GWTC-3), 
LIGO-Virgo-KAGRA collaborations obtained $H_0=68^{+8}_{-6}$ km/s/Mpc based on the $\Lambda$CDM model \cite{LIGOScientific:2021aug}.
The independent determination of the Hubble constant with GW standard sirens 
enables the potential of not only shedding light on the  the Hubble tension but also constraining other cosmological parameters \cite{Kyutoku:2016zxn}. 
There are lots of studies on the precise determination of the Hubble constant with GW standard sirens in the literature \cite{Kyutoku:2016zxn,Chen:2017rfc,Hotokezaka:2018dfi,LIGOScientific:2018gmd,LIGOScientific:2019zcs,DES:2019ccw,DES:2020nay,Cai:2016sby,Vitale:2018wlg,Mortlock:2018azx,CalderonBustillo:2020kcg,Zhu:2021bpp,Wang:2020dkc,Trott:2021fnx,Huang:2022rdg}.
There are also discussions on the uncertainties of GW standard sirens \cite{Huang:2022rdg,Chen:2020dyt}.

Due to the short arm length and various ground noises, ground-based detectors are not sensitive to GWs below 1 Hz, 
and a single detector cannot locate the source.
Space-based detectors such as the Laser Interferometer Space Antenna (LISA) \cite{Danzmann:1997hm,LISA:2017pwj}, Taiji \cite{Hu:2017mde} and TianQin \cite{TianQin:2015yph},
are sensitive to GWs in the frequency range $10^{-4}-10^{-1}$ Hz, can detect and locate mergers of distant MBBHs.
Furthermore, the network of LISA, TianQin and Taiji can significantly improve the accuracy of parameter estimation \cite{Ruan:2020smc,Zhang:2020hyx,Zhang:2020drf,Zhang:2021kkh,Gong:2021gvw,Zhang:2021wwd}.
Since the local measurement of the Hubble constant from SNe Ia data is independent of cosmological models, 
if we can use the accurate distance measurement from GWs to calibrate the luminosity distances of SNe Ia data, 
then we can use SNe Ia to determine the local Hubble constant without the problem of zero-point calibration.
The idea of using GWs as a new cosmic distance ladder for an independent calibration of distances to SNe Ia was discussed for mergers of BNS in \cite{Zhao:2017imr,Gupta:2019okl}.
Zhao and Santos used the event GW170817 to measure the absolute magnitude of SNe Ia \cite{Zhao:2017imr}.
In Ref. \cite{Gupta:2019okl}, 
the authors found that a third-generation ground-based GW detector network will measure distances with an accuracy of $\sim 0.1\%-3\%$ for BNS within $\le 300$ Mpc.
However, the calibration method with BNS as standard sirens applies to low-redshift SNe Ia only and it may miss the possible variation in the absolute magnitude with the redshift.
The calibration of distances to SNe Ia with MBBH mergers is more interesting and beneficial.
Exploring the calibration over a substantial redshift range might allow for a study of potential variation in the absolute magnitude with the redshift.
Moreover, the merger of MBBHs could also be used to calibrate Gamma-Ray Bursts at high redshifts \cite{Nissanke:2013fka,Dalal:2006qt}.
LISA will detect  MBBH mergers up to the redshift $\sim 15-20$ \cite{LISACosmologyWorkingGroup:2022jok}.
As much more SNe Ia data and GW detections with space-based GW detectors will be available in the future,
it is highly possible that MBBH merges and SNe Ia occur in the same host galaxy.
Although there are many estimates on the merger rates of MBBHs \cite{Klein:2015hvg,LISACosmologyWorkingGroup:2022jok,Sesana:2008ur,Li:2022fno,Katz:2019qlu,Ricarte:2018mzn,Mangiagli:2022niy},
there is a great uncertainty about the detection rates of MBBH mergers with LISA \cite{Sesana:2008ur,Li:2022fno,Katz:2019qlu,Ricarte:2018mzn,Mangiagli:2022niy}.
However, the Athena and LISA observatories will open the exciting possibility of truly concurrent electromagnetic and GW studies of MBBHs
\cite{Piro:2022zos}.

In this paper, we consider the possibility of re-calibrating the luminosity distances of SNe Ia by GWs from MBBH merges
and the precision of the Hubble constant determined with the re-calibrated SNe Ia data.
Even though we only use low-redshift SNe Ia data to determine the local Hubble constant so that the result is independent of cosmological models, 
the calibration of the absolute distance scale for the luminosity distances is not limited to low-redshift SNe Ia data.
We consider all possible coincidences of MBBH merges and SNe Ia to re-calibrate the luminosity
distances of SNe Ia with GWs,
these re-calibrated SNe Ia include all possible redshift ranges.
Once we solve the problem of zero-point calibration for SNe Ia data, we use low-redshift SNe Ia data to determine the local Hubble constant.

The paper is organized as follows.
In Sec. \ref{sec.2}, we use the Fisher information matrix (FIM) method to estimate the accuracy of the luminosity distance from GW observations.
In Sec. \ref{sec.3}, we discuss the accuracy of the absolute magnitude of SNe Ia calibrated by GWs.
Then we determine the local Hubble constant from the SNe Ia data in Sec. \ref{sec.4}.
The conclusion is drawn in Sec. \ref{sec.5}.

\section{The measurement of luminosity distance with space-based GW detectors}
\label{sec.2}

In terms of the polarization tensor $e^A_{ij}$ with $A=+,\times$ representing the plus and cross polarizations,
the time-domain GW signal is expressed as
\begin{equation}
\label{gwsignal}
h_{ij}(t)=\sum_{A=+,\times} e^A_{ij}h_A(t),
\end{equation}
where $i,j=1,2,3$ denote the spatial components and $t$ is the coordinate time.
The output of the GW signal in the detector $\alpha$ is
\begin{equation}\label{signal}
s_{\alpha}(t)=\sum_{A} F_{\alpha}^{A} h_{A}(t)e^{i\phi_{D}(t)}+\hat{n}_{\alpha}(t),
\end{equation}
where $F^{A}_\alpha$ is the response function, $\hat{n}_{\alpha}(t)$ is the detector noise and $\phi_D(t)$ is the Doppler phase.
The Doppler phase $\phi_{D}$ is
\begin{equation}
\label{Doppler}
\phi_{D}(t)=\frac{2 \pi f R}{c} \sin \theta \cos \left(\frac{2 \pi t}{P}-\phi-\phi_{\alpha}\right),
\end{equation}
where the distance $R$ between the earth and the sun is 1 AU, $\theta$ and $\phi$ are the angular coordinates of the GW source, 
$c$ is the speed of light, $\phi_{\alpha}$ is the detector's ecliptic longitude at $t=0$ and $P=1$ year is the rotational period.
For GWs propagating in the direction $\hat{\omega}$,
the response function $F_{\alpha}^{A}=\sum_{i, j} D_{\alpha}^{i j} e_{i j}^{A}$, where
the detector tensor $D_{\alpha}^{i j}$ is
\begin{equation}
\label{detector tensor}
D^{i j}=\frac{1}{2}\left[\hat{u}^{i} \hat{u}^{j} T(f, \hat{u} \cdot \hat{\omega})-\hat{v}^{i} \hat{v}^{j} T(f, \hat{v} \cdot \hat{\omega})\right],
\end{equation}
$\hat{u}$ and $\hat{v}$ are the unit vectors for the two arms of the interferometer,
the transfer function $T(f, \hat{u} \cdot \hat{\omega})$ for the detector is \cite{estabrook1975response,Cornish:2001qi},
\begin{equation}
\label{transfer function}
\begin{split}
T(f, \hat{u} \cdot \hat{w}) &=\frac{1}{2}\left\{\operatorname{sinc}\left[\frac{f(1-\hat{u} \cdot \hat{\omega})}{2 f^{*}}\right] \exp \left[\frac{f(3+\hat{u} \cdot \hat{\omega})}{2 i f^{*}}\right]\right.\\
&\left.+\operatorname{sinc}\left[\frac{f(1+\hat{u} \cdot \hat{\omega})}{2 f^{*}}\right] \exp \left[\frac{f(1+\hat{u} \cdot \hat{\omega})}{2 i f^{*}}\right]\right\},
\end{split}
\end{equation}
sinc($x$) = sin($x$)/$x$, and $f^{*}=c/(2\pi L)$ is the transfer frequency of the detector with the arm length $L$.

We usually work in the frequency domain,
so we Fourier transform $h_A(t)$ and $n(t)$ to $h_A(f)$ and $n(f)$.
By assuming that the noises of the detector are stationary and Gaussian,
we describe the noise with the spectral density $P_{n}(f)$,
\begin{equation}
\label{noise uncorrelated}
\left\langle n(f) n\left(f^{\prime}\right)^{*}\right\rangle=\frac{1}{2} \delta\left(f-f^{\prime}\right) P_{n}(f),
\end{equation}
where $\langle...\rangle$ denotes the ``expectation value" over many noise realizations and $n^*(f)$ is the complex conjugate of $n(f)$.
For space-based GW detectors,
the noise curve is \cite{Robson:2018ifk}
\begin{equation}\label{noise curve}
P_{n}(f)=\frac{S_{x}}{L^{2}}+\frac{2\left[1+\cos ^{2}\left(f / f^{*}\right)\right] S_{a}}{(2 \pi f)^{4} L^{2}}\left[1+(0.4 \mathrm{mHz} / f)^{2}\right],	
\end{equation}
where $S_{x}$ is the position noise and $S_{a}$ is the acceleration noise.
For LISA \cite{LISA:2017pwj}, $S_{x}=(1.5\times 10^{-11} \text{ m})^{2}\ \text{Hz}^{-1}$, 
$S_{a}=(3\times 10^{-15} \text{ m s}^{-2})^{2} \text{ Hz}^{-1}$, 
$L=2.5\times 10^{9}$ m and $f^{*}=19.09$ mHz.
For TianQin \cite{TianQin:2015yph}, $S_{x}=(10^{-12} \text{ m})^{2} \text{ Hz}^{-1}$,
$S_{a}=(10^{-15} \text{ m s}^{-2})^{2} \text{ Hz}^{-1}$,
$L=\sqrt{3}\times 10^{8}$ m and $f^{*}=0.2755$ Hz.
For Taiji \cite{Ruan:2020smc}, 
$S_{x}=(8\times 10^{-12} \text{ m})^{2} \text{ Hz}^{-1}$, 
$S_{a}=(3\times 10^{-15} \text{ m s}^{-2})^{2} \text{ Hz}^{-1}$, 
$L=3\times 10^{9}$ m and $f^{*}=15.90$ mHz.

For LISA and Taiji, we also consider the confusion noise \cite{Robson:2018ifk}
\begin{equation}\label{confusion}
\begin{split}
S_{c}(f)=& \frac{2.7 \times 10^{-45} f^{-7 / 3}}{1+0.6(f / 0.01909)^{2}} e^{-f^{0.138}-221 f \sin (521 f)} \\
& \times[1+\tanh (1680(0.00113-f))] \text{ Hz}^{-1}.
\end{split}
\end{equation}

In the frequency domain, the GW waveform $h_A(f)$ for the dominant harmonic is
\begin{equation}
\label{waveform}
\begin{split}
&h_{+}(f)= \frac{1+\cos ^{2}(\iota)}{2} \mathcal{A}(f) e^{i\Psi(f)}, \\
&h_{\times}(f)=i \cos (\iota) \mathcal{A}(f) e^{i\Psi(f)},
\end{split}
\end{equation}
where $\iota$ is the inclination angle of the orbit relative to the line of sight.
For simplicity, we consider the PhenomA waveform for a coalescing binary.
In the inspiral stage, the amplitude $\mathcal{A}$ and the phase up to the second order post-Newtonian approximation for the PhenomA waveform are \cite{Ajith:2007qp,Ruan:2019tje}
\begin{equation}
\label{amplitude}
\mathcal{A}(f) = \sqrt{\frac{5}{24}} 
\frac{\left(G \mathcal{M}_{c} / c^{3}\right)^{5 / 6} f^{-7 / 6}}{\pi^{2 / 3}\left(d_{L} / c\right)},
\end{equation}
\begin{equation}\label{phase}
\begin{split}
\Psi=& 2 \pi f t_{c}-\phi_{c}-\frac{\pi}{4}+\frac{3}{128 \eta}\left[\nu^{-5}+\left(\frac{3715}{756}+\frac{55}{9} \eta\right) \nu^{-3}\right.\\
&\left.-16 \pi \nu^{-2}+
\left(\frac{15293365}{508032}+\frac{27145}{504} \eta+\frac{3085}{72} \eta^{2}\right) \nu^{-1}\right],
\end{split}
\end{equation}
where $\nu= (\pi G M f/c^{3})^{1/3}$, 
$M=m_{1}+m_{2}$ is the total mass of the binary, 
$\mathcal{M}_{c}=(m_{1}m_{2})^{3/5}/M^{1/5}$ is the chirp mass, 
$\eta=m_{1}m_{2}/M^{2}$
is the symmetric mass ratio, the luminosity distance
\begin{equation}
\label{dl}
d_{L}(z)=c (1+z) \int_{0}^{z} \frac{d z^{\prime}}{H\left(z^{\prime}\right)}
\end{equation}
for a flat Universe, $z$ is the redshift,
the Hubble parameter 
\begin{equation}
\label{Hz}
    H(z)=H_{0}\sqrt{\Omega_{m0}(1+z)^{3}+\Omega_{\Lambda}}
\end{equation} 
for the flat $\Lambda$CDM model, the energy density for the cosmological constant $\Omega_{\Lambda}=1-\Omega_{m0}$, $\Omega_{m0}$ is the fractional matter energy density at present and $H_0$ is the Hubble constant.

\subsection{The FIM method}
\label{FIMmethod}

To use the method of match filtering to analyze signals,
we define the noise-weighted inner product for two signals $s_{1}(f)$ and $s_{2}(f)$ as
\begin{equation}\label{inner}
\left(s_{1}| s_{2}\right)=2 \int_{f_{\text {low }}}^{f_{\text {up}}} \frac{s_{1}(f) s_{2}^{*}(f)+s_{1}^{*}(f) s_{2}(f)}{P_{n}(f)} df,
\end{equation}
where  the upper cutoff frequency $f_{\text {up}}$ is chosen as the frequency $f_{\text{ISCO}}$ at the innermost stable orbit (ISCO),
\begin{equation}\label{ISCO}
f_{\text{ISCO}} = \frac{c^{3}}{6\sqrt{6}\pi GM}.
\end{equation}
Since space-based GW detectors are insensitive to GWs with frequencies below around $2 \times 10^{-5}$ Hz \cite{Baibhav:2020tma}, so we take $2 \times 10^{-5}$ Hz as the lower cutoff frequency.
For the observation of one year,
we calculate the frequency $f_{0}$ one year before the ISCO,
then we set $f_{\text {low}}=\text{max} (f_{0},2 \times 10^{-5})$.

The SNR $\rho$ for a signal $s(f)$ is
\begin{equation}\label{snr}
\rho^{2} =(s | s).	
\end{equation}
The threshold of detecting a signal is set as  $\rho \ge 8$.
For parameter estimation,
we define the FIM in the frequency domain as
\begin{equation}\label{fisher}
\Gamma_{i j}=\left.\left(\frac{\partial s(f)}{\partial \lambda_{i}} \right| \frac{\partial s(f)}{\partial \lambda_{j}}\right),
\end{equation}
where $\lambda_{i}$ is the parameter of the GW source.
The covariance matrix $\sigma_{i j}$ between the parameter errors $\Delta\lambda_i=\lambda_i-\langle\lambda_i\rangle$ and $\Delta\lambda_j$ in the large SNR limit is
\begin{equation}\label{covariance}
\sigma_{i j}=\left\langle\Delta \lambda_{i} \Delta \lambda_{j}\right\rangle \approx\left(\Gamma^{-1}\right)_{i j}.
\end{equation}
The root mean square error of the parameter $\lambda_i$
is
\begin{equation}
\label{peeq1}
\sigma_{i}=\sqrt{\sigma_{ii}}\approx \sqrt{\left(\Gamma^{-1}\right)_{ii}}.
\end{equation}
In this way, the error of the luminosity distance can be estimated from the FIM $\Gamma_{ij}$.

For a network with $n$ detectors, the SNR and FIM are $\rho^{2}=\Sigma_{\alpha=1}^{n} \rho_{\alpha}^{2}$ and $\Gamma_{i j}=\sum_{\alpha=1}^{n} \Gamma_{i j}^{\alpha}$, respectively.

\subsection{The accuracy of the luminosity distance}
\label{select}

We consider a nonspinning MBBH with 9 parameters:
the chirp mass $\mathcal{M}_{c}$, the symmetric mass ratio $\eta$,
the luminosity distance $d_{L}$, the sky location $(\theta,\phi)$,
the inclination angle $\iota$, the polarization angle $\psi$ and
the coalescence phase $\phi_{c}$ at the coalescence time $t_{c}$,
i.e., $\bm{\lambda}=(\mathcal{M}_{c}, \eta, d_{L}, \theta, \phi, \iota, \psi, t_{c}, \phi_{c})$.
For equal-mass MBBHs we considered, $\eta=0.25$.
The parameters $\iota$, $\psi$, $\phi_{c}$, $t_{c}$
are chosen randomly in the following range:
$\iota \in [0, \pi]$, $\psi \in [0, 2\pi]$, $\phi_{c} \in [0, 2\pi]$,
and $t_{c} \in [0, 1]$ in the unit of year.
The angular uncertainty of the sky localization is evaluated as
\begin{equation}
	\Delta \Omega_s=2\pi\left|\sin\theta\right|
	\sqrt{\sigma_{\theta\theta}\sigma_{\phi\phi}-\sigma^2_{\theta\phi}}.
\end{equation}

For each GW source, we assume that we can find an SN Ia which is in the same host galaxy as the GW source, 
so we use the same parameters ($\theta$, $\phi$, $z$) from the SNe Ia data for the GW source.
In this paper, we use the Pantheon sample of SNe Ia data \cite{Pan-STARRS1:2017jku}.
The Pantheon sample compiles 1048 SNe Ia data, covering the redshift range $0.01 < z < 2.26$.
We use the $\Lambda$CDM model to calculate the luminosity distance $d_{L}$ from the redshift $z$.
The cosmological parameters are chosen as the Planck 2018 results: $H_{0} = 67.27$ km/s/Mpc, and $\Omega_{m0}=0.3166$ \cite{Planck:2018vyg}.

\begin{figure}
\centering
\includegraphics[width=0.85\columnwidth]{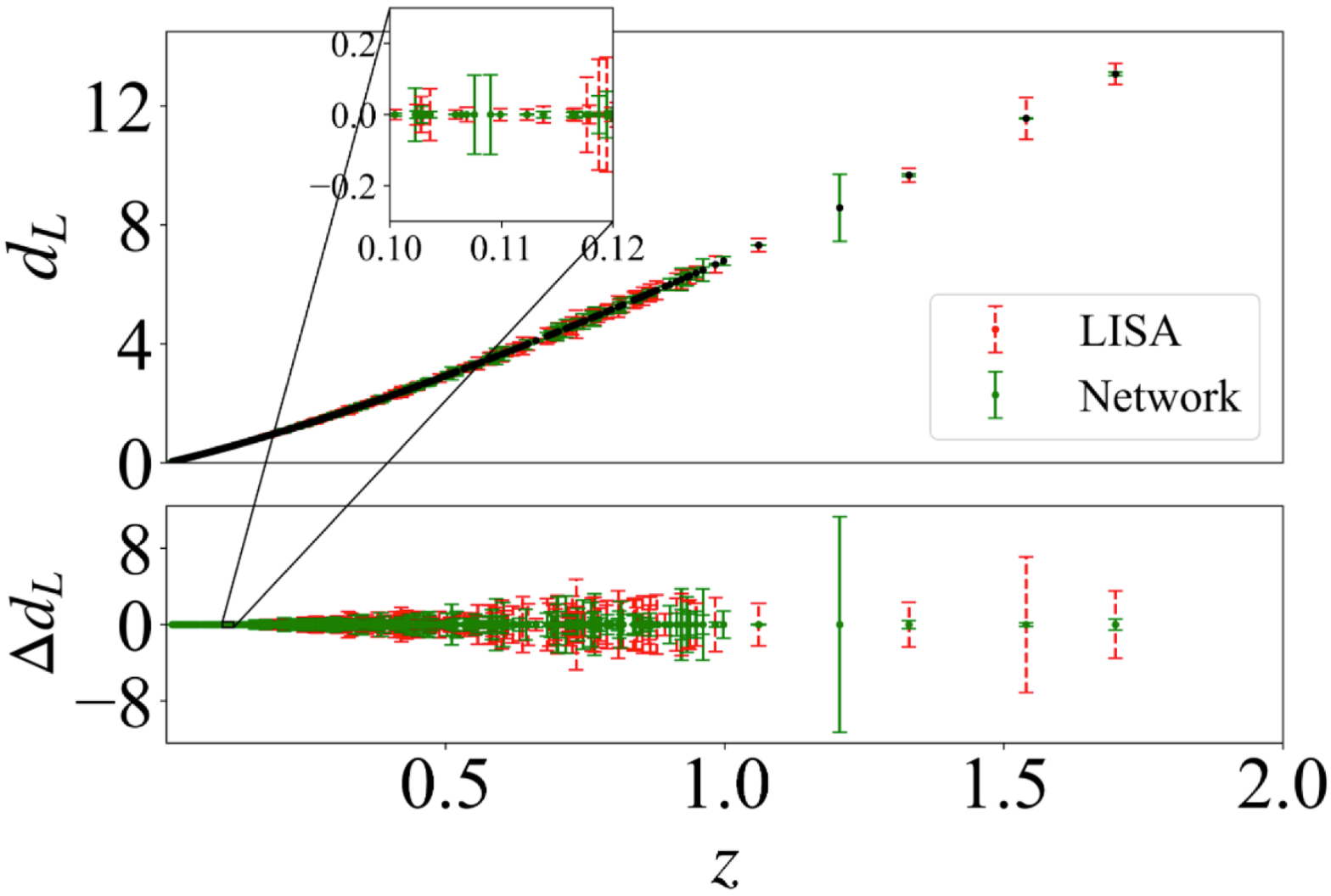}
\caption{The $1\sigma$ errors of the luminosity distance with LISA and the LISA-Taiji-TianQin network for the pop model.
In the top panel, the luminosity distances along with their estimated $1\sigma$ errors in the unit of 1 Gpc are shown.  
In the bottom panel, we show $\Delta d_L$ in the unit of 100  Mpc, 
the red dashed lines represent the estimated $1\sigma$ error bar with LISA,
and green solid lines represent the estimated $1\sigma$ error bar with the LISA-Taiji-TianQin network.}
\label{dlerrorfig}
\end{figure}

MBHs are assumed to form from seed BHs through merger and gas accretion \cite{Volonteri:2007ax, Linder:2002et}.
For MBBHs, following Ref. \cite{Klein:2015hvg, Wang:2020dkc}, 
we consider the three widely accepted population models: pop III, Q3d, and Q3nod.
The pop III model assumes that the MBH seeds are the remnants of population III stars with initial masses centered around $300 M_{\odot}$ at $z \approx 15 \sim 20$.
Both the Q3d and Q3nod models assume MBHs seed from the collapse of protogalactic disks and already have  masses around $10^5 M_{\odot}$ at the redshift $z \approx 15 \sim 20$, 
but the former model takes into account the delays between the formation of MBHs and galaxy mergers, 
while the latter model does not.
The distributions of the redshift and mass of MBBHs for the three seed models can be found in the Fig. 1 of Ref. \cite{Wang:2020dkc}.
From the distributions of the redshift and the total mass of MBBHs
based on the three population models,
we generate a set of MBBHs with some chirp mass $\mathcal{M}_{c}$ and redshift determined by the distribution.
Using GWs from these MBBH mergers, we estimate the luminosity distance error and the angular resolution with the FIM method and the results are shown in Fig. \ref{dlerrorfig} and Table \ref{tab1}.
The results for the three models are similar, 
so we only show the results obtained with the pop III model in Fig. \ref{dlerrorfig} and all the figures in the following discussions.
The results are consistent with those in Ref. \cite{Gong:2021gvw,Zhang:2021wwd,Crowder:2005nr,Ruan:2020smc,Ruan:2019tje}.
For the same detection threshold $\rho \ge 8$, the LISA-Taiji-TianQin network can detect some GW signals that can not be detected by LISA alone,
this is the reason why some results with the network only appear in Fig. \ref{dlerrorfig}.

\begin{table}[htbp]
\centering
\begin{tabular}{*{7}{c}}
  \toprule
  & \multicolumn{3}{c}{Luminosity Distance} & \multicolumn{3}{c}{Angular Resolution (deg$^{2}$)} \\
  \cmidrule(r){2-4}\cmidrule(r){5-7}
  Models & pop & Q3d  & Q3nod  & pop & Q3d & Q3nod \\
  \midrule
   LISA     &$2.08\%$  &$1.58\%$   &$1.28\%$  
	        &$1.5\times 10^{-1}$   &$3.8\times 10^{-1}$ &$2.1\times 10^{-1}$ 
  \\ 
  Network   &$0.25\%$  &$0.03\%$. &$0.06\%$  
            &$4.8\times 10^{-4}$   &$1.7\times 10^{-5}$ &$5.1\times 10^{-5}$         
  \\ 
  \bottomrule
\end{tabular}
\caption{The median values of the  relative error of the luminosity distance and the angular resolution with LISA and the LISA-Taiji-TianQin network for different population models.}
\label{tab1}
\end{table}

From Fig. \ref{dlerrorfig} and Table \ref{tab1}, we see that the median value of the relative error of the luminosity distance is larger than $1\%$ and the median value of the angular resolution is bigger than $0.1$ deg$^2$ with LISA.
The Q3nod model gives a better constraint on the luminosity distance at redshift $z\lesssim 1.5$,
but the pop model gives a better constraint on the angular resolution.
To improve the accuracy of the distance measurement and the sky localization of the source,
we use the network of LISA, Taiji and TianQin (LISA-Taiji-TianQin) to make parameter estimation \cite{Zhang:2020hyx,Zhang:2020drf,Zhang:2021kkh,Gong:2021gvw,Zhang:2021wwd}
and the results are shown in Fig. \ref{dlerrorfig} and Table \ref{tab1}. 
With the LISA-Taiji-TianQin network, the Q3d model can give the luminosity distance at the precision level of $0.03\%$ and improve the angular resolution to reach $1.7\times 10^{-5}$ deg$^2$.
Therefore, the network improves the accuracy of the sky localization and the luminosity distance than that with LISA alone by several orders of magnitude.
Take the median values obtained with the Q3nod model and the LISA-Taiji-TianQin network: 
$d_L\sim 1300$ Mpc, $\Delta d_L\sim 0.8$ Mpc, $\Delta \Omega_s\sim 5.1\times 10^{-5}$ deg$^2$,
we estimate the uncertainty of the volume that the source is located in as $\Delta V\sim  6.7\times 10^{-8}$ Gpc$^3$. 
Since the number density of galaxies is about $3\times 10^6$ Gpc$^{-3}$ \cite{Gupta:2019okl}, 
the error of the localized volume will contain no more than one field galaxy.
Once the MBBH is located within one galaxy, then the host galaxy can be identified and we can determine whether there is a SN Ia occurred in the same galaxy. 
If MBBH mergers and SNe Ia occur in the same host galaxy,
then we can calibrate standard candles with standard sirens.

\section{The calibration error of the absolute magnitude}
\label{sec.3}

In this section, we assume that an MBBH merger and an SN Ia are in the same host galaxy
so that we can use the luminosity distance of the MBBH with GW measurement to calibrate the SN Ia.
At the redshift $z$, the apparent magnitude $m_{B}(z)$ of an SN Ia is
\begin{equation}
\label{mB}
m_{B}(z)=5 \log _{10}\left[\frac{d_{L}(z)}{\text{Mpc}}\right]+25+M_{B},
\end{equation}
where $M_{B}$ is the absolute magnitude.
The error in the estimation of the absolute magnitude (calibration error) mainly comes from
the measurement uncertainties of the apparent magnitude $\sigma_{m_B}$ and the luminosity distance $\sigma_{d_L}$,
so the error of the absolute magnitude is
\begin{equation}
\label{sigmb}
\sigma_{M_B}=\sqrt{(\sigma_{m_{B}})^{2}+\left(\frac{5 \sigma_{d_{L}}}{\ln 10\, d_{L}}\right)^{2}}.
\end{equation}
For convenience, we define
\begin{equation}
\label{mbbydl}
\sigma_{*}= \frac{5\sigma_{d_L}}{\ln 10\, d_{L}}.	
\end{equation}
The error of the luminosity distance can be large at some locations.
To reduce the calibration error of the absolute magnitude,
we discard those GW events with the signal-to-noise ratio $\rho<8$ and $\sigma_{*}> \sigma_{m_{B}}$ detected by LISA. 
With this cutoff, we are left with 679 SNe Ia data for the pop model, 743 SNe Ia data for the Q3d model and 804 SNe Ia data for the Q3nod model.
Note that we already applied this cutoff in Fig. \ref{dlerrorfig}.

\begin{figure}[htbp]
	\centering
	\includegraphics[width=0.45\columnwidth]{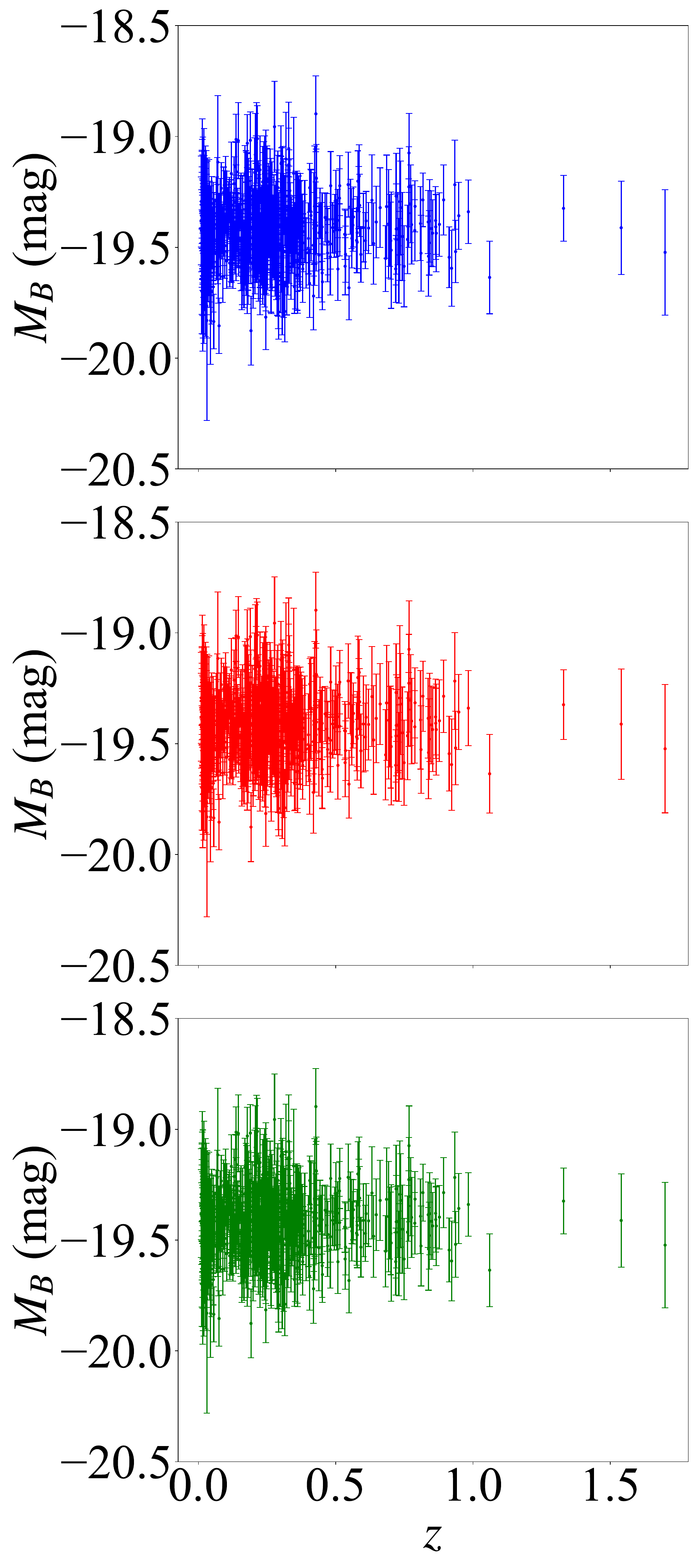}
	\caption{The absolute magnitude $M_{B}$ with $1\sigma$ uncertainty calibrated by GWs with the pop model.
			The top panel shows the observed apparent magnitude, i.e., no error of $d_{L}$ is included. In the middle and bottom panels, we include the errors of $d_{L}$ measured by LISA and the LISA-Taiji-TianQin network, respectively.
			}
	\label{MBerrorbarsfig}
\end{figure}

With the estimated luminosity and the observed apparent magnitude for each SN Ia, we calculate $M_B$ and $\sigma_{M_{B}}$ for each SN Ia and the results are shown in Fig. \ref{MBerrorbarsfig}.
We also summarize the median, mean and minimum values of $\sigma_{m_{B}}$ and $\sigma_{M_{B}}$ for all the SNe Ia data in Table \ref{MBtable}.
From Table \ref{MBtable}, we see that the error of the luminosity distance accounts for less than $10 \%$ error of the absolute magnitude.
In particular, for the LISA-Taiji-TianQin network, $\sigma_{M_{B}}$ is almost the same as $\sigma_{m_{B}}$, so the contribution of $\sigma_{*}$ to $\sigma_{M_{B}}$ is almost negligible.
Fig. \ref{MBerrorbarsfig} also shows that the calibration error is mainly from the measurement uncertainty of the apparent magnitude.

\begin{table}[htbp]
\centering
\begin{tabular}{*{10}{c}}
  \toprule
  & \multicolumn{3}{c}{Median Value} & \multicolumn{3}{c}{Mean Value} & \multicolumn{3}{c}{Minimum Value} \\
  \cmidrule(r){2-4}\cmidrule(r){5-7}\cmidrule(r){8-10}
  Models & pop & Q3d  & Q3nod  & pop & Q3d & Q3nod & pop & Q3d  & Q3nod \\
  \midrule
   $\sigma_{m_{B}}$   &$0.1353$   &$0.1360$    &$0.1354$    &$0.1396$   &$0.1401$   &$0.1403$   & $0.0854$    &$0.0939$    & $0.0854$        \\
   $\sigma_{M_{B}}$ (LISA)        &$0.1428$   &$0.1422$    &$0.1415$    &$0.1485$   &$0.1488$   &$0.1467$   & $0.0962$    &$0.0981$    & $0.0963$         \\
   $\sigma_{M_{B}}$ (Network)     &$0.1363$   &$0.1360$    &$0.1356$    &$0.1407$   &$0.1410$   &$0.1403$   & $0.0854$    &$0.0939$    & $0.0854$         \\
  \bottomrule
\end{tabular}
\caption{The median, mean and minimum values of $\sigma_{m_{B}}$ and $\sigma_{M_{B}}$. $\sigma_{M_{B}}$ (LISA) means the result for $\sigma_{M_{B}}$ with LISA,
and $\sigma_{M_{B}}$ (Network) means the result for $\sigma_{M_{B}}$ with the  LISA-Taiji-TianQin network.}
\label{MBtable}

\end{table}

\begin{figure}[htbp]
	\centering
	\includegraphics[width=0.65\columnwidth]{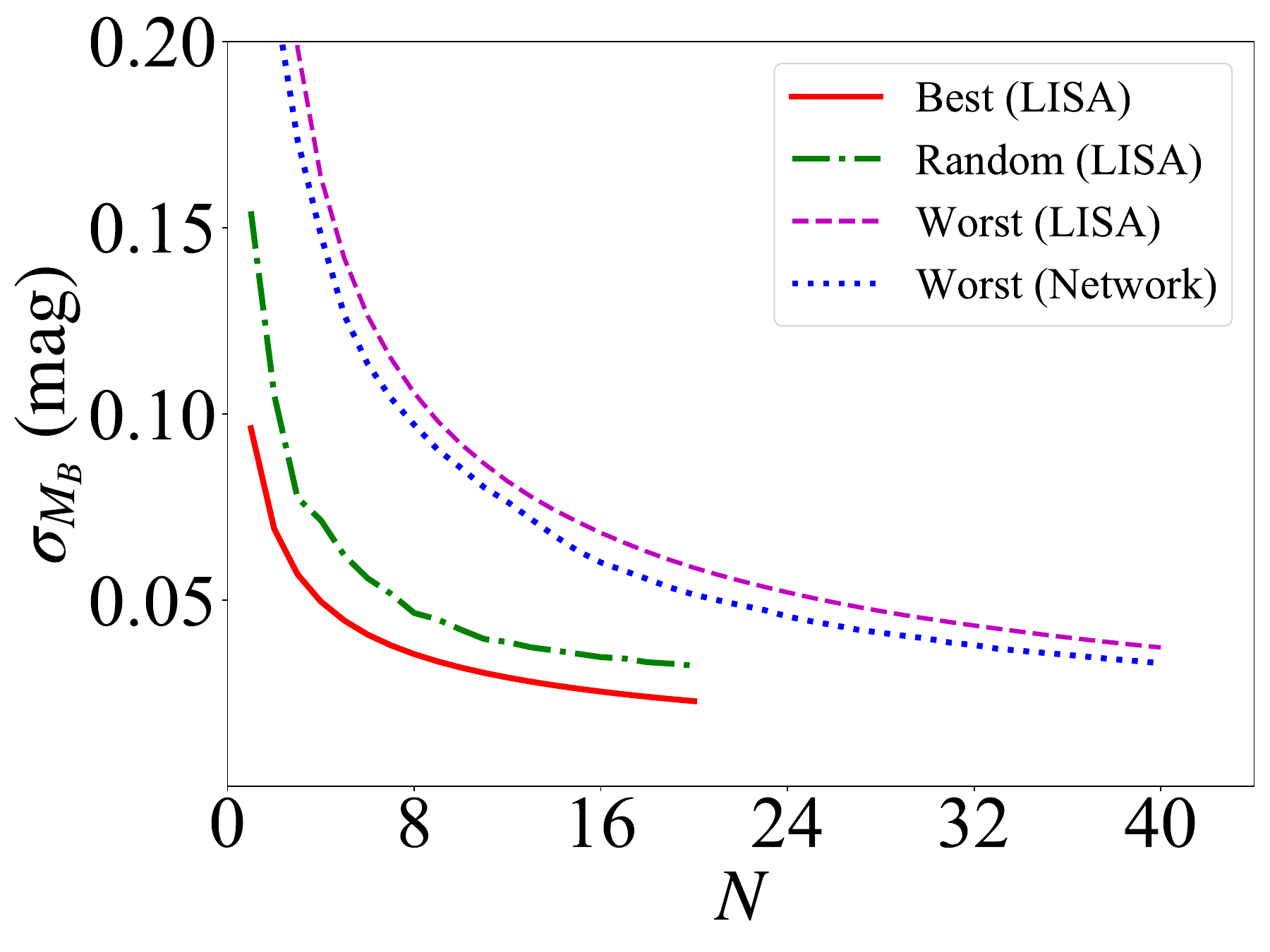}
	\caption{The dependence of $\sigma_{M_{B}}$ on the number of calibrators $N$ for the pop model.
	The red solid line and the green dash-dot line represent the estimated 1$\sigma$ error of $M_{B}$ for the best scenario and the random scenario with LISA,
	the magenta dashed line represents the estimated 1$\sigma$ error of $M_{B}$ for the worst scenario with LISA,
	the blue dotted line represents the estimated 1$\sigma$ error of $M_{B}$ for the worst scenario with the LISA-Taiji-TianQin network.
           }
	\label{multipleMBfig}
\end{figure}

The above discussion assumes that we have only one calibrator.
Now we consider the calibrations of more than one SN Ia.
In other words, we assume that we can locate $N$ pairs of MBBH mergers and SNe Ia that each pair is in the same host galaxy,
so that we have $N$ GW-calibrated SNe Ia to reduce statistical error.
We discuss three cases, the best scenario considers those SNe Ia with the smallest measurement error on the apparent magnitude, 
the worst scenario considers those SNe Ia with the biggest $\sigma_{m_B}$,
and the random scenario selects SNe Ia randomly.
To constrain $M_{B}$ with $N$ calibrators, we minimize
\begin{equation}\label{chi2}
\chi^{2}=\sum_{i=1}^{N}\left[\frac{m_{B}^{i}-m_{B}\left(d_{L}^{i}, M_{B}\right)}{\sigma^i}\right]^{2}
\end{equation}
with iminuit \cite{iminuit},
and the results of $\sigma_{M_{B}}$ versus the number $N$ are shown in Fig. \ref{multipleMBfig}.
Here $m_{B}^{i}$ is the observed apparent magnitude for the SN Ia at the redshift $z_i$,
$m_{B}(d_{L}^{i}, M_{B})$ is obtained with Eq. \eqref{mB} and
 $\sigma^i$ is
\begin{equation}
\sigma^i=\sqrt{(\sigma_{m_{B}}^i)^{2}+(\sigma_*^i)^2}.
\end{equation}

From Fig. \ref{multipleMBfig}, we see that the error of $M_{B}$ decreases as the number of calibrators increases.
Due to the observational limit set by $\sigma_{m_B}$, the improvement on $\sigma_{M_B}$ by larger $N$ is not significant once $N$ reaches a certain value,
and the results from LISA alone and the LISA-Taiji-TianQin network are similar for the best and random scenarios.
For the best scenario, with 10 calibrators, $\sigma_{M_B}$ can reach $0.03$ mag for all three population models; 
If $N=20$, $\sigma_{M_B}$ can reach $0.023$ mag for all three population models.
For $N=10$, the highest redshift in SNe Ia data is $z=(0.30, 0.37, 0.30)$ with the model (pop, Q3d, Q3nod) for MBBHs.
For $N=20$, the highest redshift in SNe Ia data is 
$z=(0.30, 0.37, 0.37)$ with the model (pop, Q3d, Q3nod) for MBBHs.
For the random scenario, with 10 calibrators, $\sigma_{M_B}$ can reach $0.04$ mag for all three population models;
If $N=20$, $\sigma_{M_B}$ can reach $0.028$ mag for the Q3nod model.
For $N=10$, the highest redshift in SNe Ia data is $z=(0.53, 0.70, 1.33)$ with the model (pop, Q3d, Q3nod) for MBBHs.
For $N=20$, the highest redshift in SNe Ia data is $z=(0.78, 0.70, 1.33)$ with the model (pop, Q3d, Q3nod) for MBBHs.
For the worst scenario, with 20 calibrators, $\sigma_{M_B}$ can reach $0.05$ mag for all three population models;
If $N=40$, $\sigma_{M_B}$ can reach $0.034$ mag for all three population models.
For $N=20$ or $N=40$, the highest redshift is $z=1.7$ for all three models.
The uncertainty $\sigma_{M_B}$ with the LISA-Taiji-TianQin network is better than that with LISA alone in the worst scenario case.
Even though the LISA-Taiji-TianQin network does not help much on the reduction of $\sigma_{M_B}$ for the best and random scenarios,
the much more accurate localization of the GW source with the network may be helpful in finding a companion SN Ia.

\section{The uncertainty of Hubble constant}
\label{sec.4}

In the last section, we discussed the calibrations of the Pantheon sample of SNe Ia data by GWs.
Now we can use the calibrated SNe Ia data to measure the Hubble constant $H_0$.
Since the calibration of the luminosity distance by GWs involves only one-step distance ladder, 
the measured Hubble constant can overcome the the problem from electromagnetic distance ladder.
To avoid the dependence of cosmological models, we use the kinematic $d_L-z$ relation from Taylor expansion \cite{Gong:2004sd},
\begin{equation}\label{dl*}
d_{L}(z)=\frac{c z}{H_{0}}\left[1+\frac{\left(1-q_{0}\right) z}{2}+O\left(z^{2}\right)\right],
\end{equation}
to fit low-redshift SNe Ia data, where $q_{0}$ is the deceleration parameter. 
Following Ref. \cite{Riess:2016jrr}, to avoid the possibility of a coherent flow in the more local volume,
we use 237 SNe Ia in the redshift range $0.023< z < 0.15$ to constrain the Hubble constant $H_0$ with the cosmographic expansion \eqref{dl*}.
As discussed in \cite{Camarena:2019moy},
the minimum cutoff of $z$ is large enough to reduce the impact of cosmic variance,
and the maximum $z$ is small enough to avoid the dependence on cosmological models.

Now we determine cosmological parameters $H_0$ and $q_0$ by marginalizing over $M_{B}$ with the Bayesian analysis,
\begin{gather}
\label{posterior}
f\left(H_{0},q_{0} | \text{SN}\right)=\int d M_{B} f\left(H_{0},q_{0}, M_{B} | \text{SN}\right),\\
\label{posterior1}
f(H_{0},q_{0}, M_{B} | \text{SN})=\frac{f(H_{0}) f(q_{0}) f(M_{B})\mathcal{L}(\text{SN} | H_{0}, q_{0}, M_{B})}{\mathcal{E}},
\end{gather}
where $f(H_{0})$, $f(q_{0})$ and $f(M_{B})$ are the prior distributions of $H_{0}$, $q_{0}$ and $M_{B}$, respectively,
$f(M_B)$ is a Gaussian distribution with
the mean $M_B$ and the $1\sigma$ error $\sigma_{M_B}$ given in the last section,
$\mathcal{L}$ is the likelihood, $\mathcal{E}$ is the evidence,
and SN stands for the given SNe Ia data in the redshift range $0.023\le z\le 0.15$ \cite{Pan-STARRS1:2017jku}.
The likelihood $\mathcal{L}$ is
\begin{equation}\label{likelihood}
\mathcal{L}(\text{SN} \mid H_{0}, q_{0}, M_{B})=|2 \pi \Sigma|^{-1 / 2} e^{-\frac{1}{2} \chi^{2}(H_{0}, q_{0}, M_{B})},
\end{equation}
where $\chi^{2}$ is
\begin{equation}\label{chi2*}
\chi^{2}=\left[m_{B}^{i}-m_{B}\left(z_{i}\right)\right] \Sigma_{i j}^{-1}\left[m_{B}^{j}-m_{B}\left(z_{j}\right)\right],
\end{equation}
$\Sigma$ is the covariance matrix of the 237 SNe Ia data,
and $m_{B}(z_{i})$ is the predicted apparent magnitude at the redshift $z_i$ from Eqs. \eqref{mB} and \eqref{dl*}.

\begin{figure}[htbp]
	\centering
	\includegraphics[width=0.65\columnwidth]{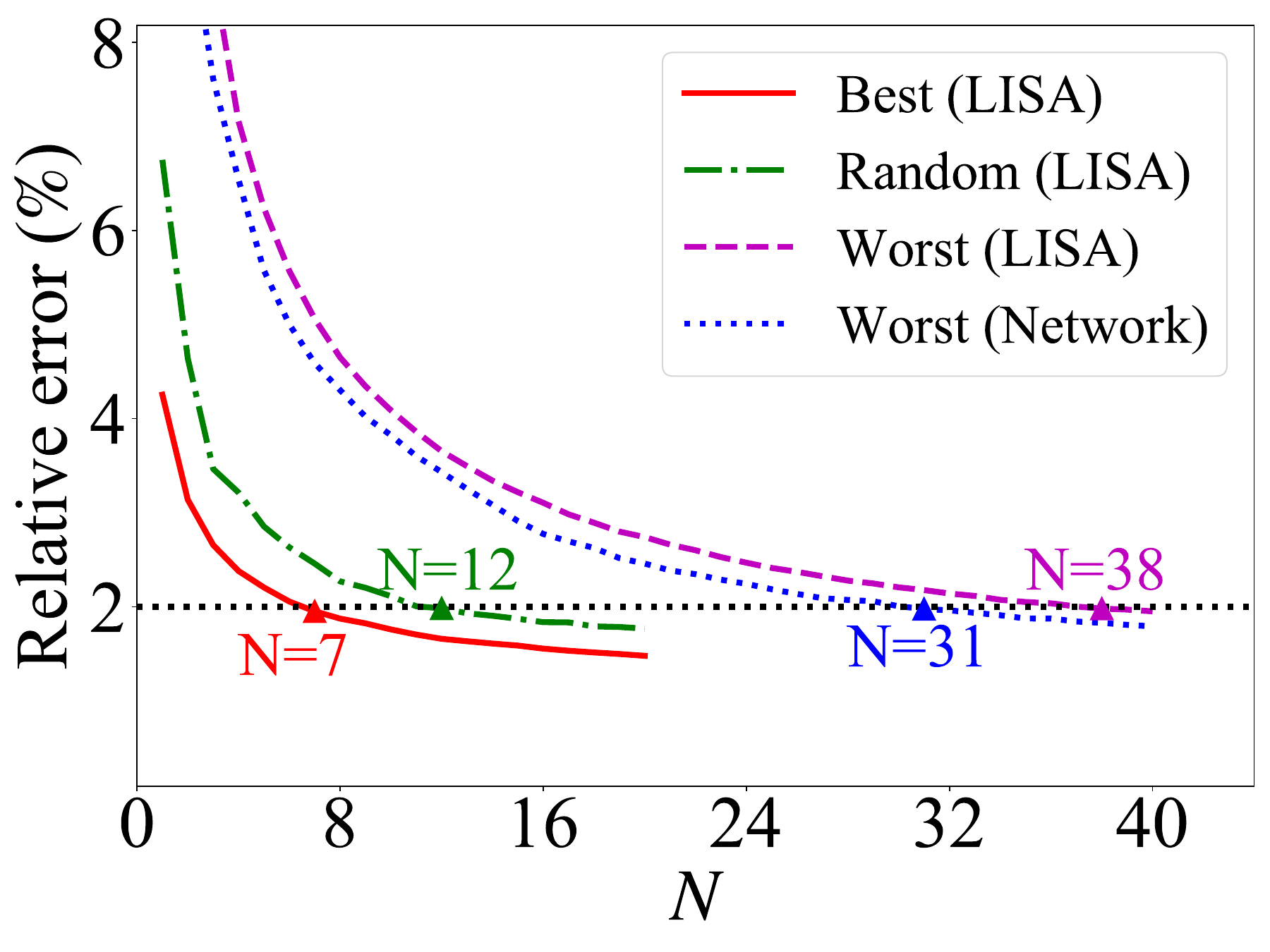}
	\caption{The relative error of $H_{0}$ with the pop model.
	The triangle represents the smallest number of calibrators $N$ needed for the relative error reaching below 2\%.
	The red solid line and the green dash-dot line represent the constrained relative error of $H_{0}$ for the best scenario and the random scenario with LISA,
	the magenta dashed line represents the constrained relative error of $H_{0}$ for the worst scenario with LISA,
	the blue dotted line represents the constrained relative error of $H_{0}$ for the worst scenario with the LISA-Taiji-TianQin network.
			}
	\label{multipleH0fig}
\end{figure}

For the best scenario, the relative error of $H_{0}$ can be less than 2\% for the three models with 7 calibrators;
If $N = 20$, the relative error of $H_{0}$ can reach 1.6\% for all three models.
The results are almost the same either with LISA alone or with the LISA-Taiji-TianQin network for all three scenarios.
For the random scenario, the relative error of $H_{0}$ can reach below 2\% with 12, 14, and 11 calibrators for the pop, Q3d, and Q3nod models, respectively;
If $N = 20$, $\sigma_{H_{0}}/H_{0}$ can be less than  1.9\% for all three models.
For the worst scenario, the relative error of $H_{0}$ can reach below 2\% with 31, 32, and 32 calibrators for the pop, Q3d, and Q3nod models, respectively;
If $N = 40$, $\sigma_{H_{0}}/H_{0}$ can be less than  1.99\% for all three models.
These results are shown in Fig. \ref{multipleH0fig}.
The results tell us that we can get a better than 2\% determination of the local value of the Hubble constant from SNe Ia in the redshift range $0.023\le z\le 0.15$ in a model independent way by calibrating the luminosity distances of about 10 SNe Ia with GWs. 
Due to the measurement uncertainty of the apparent magnitude for SNe Ia,
more calibrated SNe Ia can hardly reduce the relative error of $H_0$ further.
Since the luminosity distances of MBBHs were simulated with the flat $\Lambda$CDM model,
the central value of $H_0$ obtained here may not be trusted,
but the estimated error of $H_0$ is independent of the model.   
Once the observations of GWs from MBBHs with space-based GW detectors are available,
the method presented here can determine the local value of $H_0$ with better than 2\% precision.
However, the relative error of deceleration parameter $q_{0}$ is around 30\%.

The above simulation is based on the flat $\Lambda$CDM model with $H_{0} = 67.27$ km/s/Mpc.
To investigate the impact of the choice of the value of cosmological parameters,
we also did the simulation with the cosmological parameters
$H_{0} = 73.00$ km/s/Mpc and $\Omega_{m0}=0.3166$ \cite{Riess:2020fzl},
and we find that the results are similar.
For the best scenario, the relative error of $H_{0}$ can be less than $2\%$ with 7 calibrators by LISA or the LISA-Taiji-TianQin network.
For the random scenario, the relative error of $H_{0}$ can reach below $2\%$ with 13 calibrators by LISA.
For the worst scenario, the relative error of $H_{0}$ can reach below $2\%$ with 38 calibrators by LISA.
If we use the LISA-Taiji-TianQin network, the number of calibrators needed to reach $2\%$ accuracy for the random and worst scenarios is 12 and 32, respectively.
Therefore, the model independent determination of the local Hubble constant from SNe Ia data calibrated by GWs can shed light on the Hubble tension.

\begin{figure}[htbp]
	\centering
	\includegraphics[width=0.65\columnwidth]{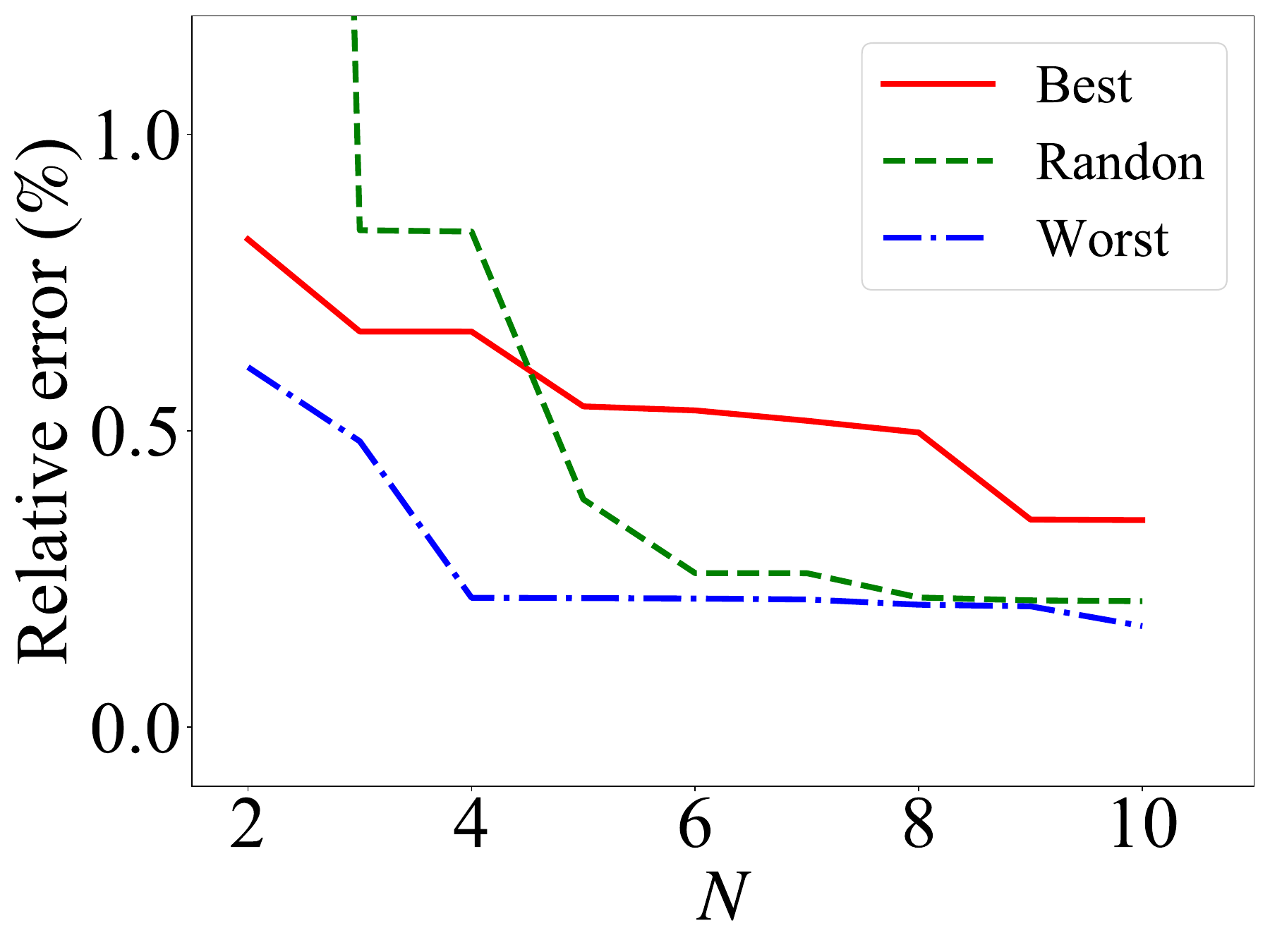}
	\caption{The relative error of $H_{0}$ determined from $N$ GW standard sirens with LISA for the pop model.
	The red solid line, the green dash-dot line, and the blue dotted line represent the constrained 1$\sigma$ relative error of $H_{0}$ for the best scenario, the random scenario, and the worst scenario, respectively.
			}
	\label{sirenfig}
\end{figure}

For comparison, we also consider those GWs which calibrate SNe Ia as standard sirens to constrain the Hubble constant.
Since the redshift of MBBHs is as large as $z\sim 0.3$ for the best scenario, $z\sim 1.3$ for the random scenario, and $z\sim 1.7$ for the worst scenario,
we cannot use the cosmographic expansion \eqref{dl*} and a cosmological model must be invoked.
For simplicity, we consider the constraint on the Hubble constant from the standard siren based on the $\Lambda$CDM model.
In Fig. \ref{sirenfig}, we show the relative error of $H_0$ determined from $N$ GW standard sirens with LISA.
The results show that the relative error can reach below 1\% with $N\gtrsim 4$ for all scenarios.
The result is consistent with that in Ref. \cite{Tamanini:2016zlh,Jin:2023sfc}.
For the LISA-Taiji-TianQin network, the relative error of $H_0$  is less than 0.1\%.
As discussed above, the results from GWs as standard sirens depend on cosmological models
even though the relative error is much smaller.

After learning that at least 7 SNe Ia with their luminosity distances calibrated by GWs are needed to reach a 2\% determination of the local Hubble constant,
we can now assess whether it will be possible to be realized within this the next decade of the operation of space-based detectors.
According to \cite{Barausse:2012fy}, the galaxies number density is $\approx 2 \times 10^7$ Gpc$^{-3}$,
so there are $8.2 \times 10^{10}$ galaxies below redshift $z=2$ ($d_L \approx 16 $ Gpc by $\Lambda$CDM model with $H_{0} = 67.27$ km/s/Mpc and $\Omega_{m0}=0.3166$),
and let us consider that MBBHs and SNe Ia uniformly distributed in the co-moving volume between redshift z of 0 and 2.
The estimate of SN Ia rate in redshift $z = 1$ is roughly $1.2 \times 10^5$ Gpc$^{-3}$ yr$^{-1}$ \cite{Li:2010kd}, 
which represents the SN Ia density at reshift [0, 2].
The estimate of MBBH rate in redshift range [0, 2] is roughly $2 \times 10^{-3}$ Gpc$^{-3}$ yr$^{-1}$ \cite{Klein:2015hvg} that there are 8 MBBHs below redshift $z=2$ per yr.
To ensure that these signals can be recognized by the LISA-Taiji-TianQin network, we simulated 3600 GW signals, where the selection of redshift and mass are according to \cite{Klein:2015hvg}, and other parameters are randomly chosen.
We found that 2824 mergers are $\rho > 8$ with the detector network. 
In other words, about 3/4 of the MBBH GW signals within the redshift [0, 2] can be detected by the LISA-Taiji-TianQin network.
Hence, SN Ia and MBBH singly occur in a galaxy roughly about once every 170 yr and $1.25 \times 10^{10}$ yr.
Thus, the odds of both SN Ia and MBBH occurring in a single galaxy over 10 years are approximately 1 in $2.1 \times 10^{10}$ per galaxy.
So as a rough estimate, we can observe 3.7 calibrators in a decade of space-based detectors.
For a longer period of detection, we can detect 7 and 30 calibrators in 14 years and 30 years.

\section{Conclusion}
\label{sec.5}

The main problem of the model independent determination of the local Hubble constant from SNe Ia is the absolute calibration of the peak brightness for SNe Ia.
The observations of GWs as one-step standard sirens can be used to calibrate the luminosity distances of SNe Ia
if an SN Ia and an MBBH merger occur in the same host galaxy.
If one SN Ia is calibrated with a GW standard siren,
we find that the measurement error of the luminosity distance with LISA accounts for less than $10 \%$ error of the absolute magnitude.
Furthermore, the contribution of the measurement error of the luminosity distance to $\sigma_{M_{B}}$ is almost negligible for the LISA-Taiji-TianQin network.
We conclude that the calibration error for SNe Ia is mainly from the measurement uncertainty of the apparent magnitude.

For $N$ calibrators, we discussed three cases, the best-case scenario assumes that $N$ SNe Ia with the smallest measurement error on the apparent magnitude and MBBH mergers occur in the same host galaxy,
the worst-case scenario assumes that $N$ SNe Ia with the biggest $\sigma_{m_B}$ and MBBH mergers occur in the same host galaxy,
and the random-case scenario assumes that $N$  randomly selected SNe Ia and MBBH mergers occur in the same host galaxy.
For each case, the measured luminosity distances are used to calibrate the absolute magnitude of $N$ SNe Ia.
For the best-case scenario, $\sigma_{M_B}$ can reach $0.023$ mag for all three population models.
The uncertainty of the absolute magnitude can be as small as $0.034$ mag even for the worst-case scenario.
Note that the redshift of the calibrated SNe Ia is not limited to be small and it can be arbitrarily large.

After re-calibrating the absolute magnitude of the Pantheon SNe Ia data,
we use 237 SNe Ia in the redshift range $0.023< z < 0.15$ to constrain the local Hubble constant.
Note that for the calibration, we are not limited to the 237 SNe Ia in the redshift range $0.023< z < 0.15$, 
we considered all possible coincident SNe Ia and MBBH mergers to calibrate the whole Pantheon sample of SNe Ia data.
For the best-case scenario, the relative error of $H_{0}$ can be less than 2\% for the three population models with 7 calibrators.
For the random-case scenario, the relative error of $H_{0}$ can reach below 2\% with 12, 14, and 11 calibrators for the pop, Q3d, and Q3nod models, respectively.
For the worst-case scenario, the relative error of $H_{0}$ can reach below 2\% with 31, 32, and 32 calibrators for the pop, Q3d, and Q3nod models, respectively.
The uncertainty of the local Hubble constant can be reduced a little bit with more number of calibrators,
but the reduction of the uncertainty is insignificant.
If we use those GWs that calibrate the luminosity distance of SNe Ia as standard sirens to determine the Hubble constant,
we can get a less than 1\% precision with LISA and less than 0.1\% precision with the LISA-Taiji-TianQin network.
However, the results based on standard sirens depend on cosmological models.
Subtleties may arise if we consider the relative positions of SNe Ia and the host galaxy of the MBBH mergers, and the peculiar velocity of the host galaxy.

We conclude that at least 7 SNe Ia with their luminosity distances calibrated by GWs are needed to reach a 2\% determination of the local Hubble constant.
The value of the local Hubble constant is free from the problems of zero-point calibration and model dependence.
Therefore, the model independent determination of the local Hubble constant from SNe Ia data calibrated by GWs can shed light on the Hubble tension.

\end{CJK*}
\begin{acknowledgements}
This work is supported by the National Key Research and Development Program of China under Grant No. 2020YFC2201504 and
the National Natural Science Foundation of
China under Grant No. 11875136.
The numerical computations were performed at the public computing service platform provided by the Network and Computing Center of HUST.
\end{acknowledgements}


\end{document}